\pdfoutput=1

\documentclass[12pt,preprint]{aastex}
\usepackage{apjfonts}

\usepackage{natbib} 
\usepackage{amsmath}
\usepackage{amssymb}
\usepackage{subfigure}
\usepackage{url}
\usepackage{CJK}

\newcommand{\rmd}{ {\ \mathrm d} }
\renewcommand{\vec}[1]{ {\mathbf #1} }

\newcommand{\grad}{ {\bf \nabla } }

\newcommand{\Eq}{{Equation}}

\newcommand{\Fig}{{Figure}}

\newcommand{\dive}{\nabla\cdot}
\newcommand{\divB}{\nabla\cdot\mathbf{B}}

\graphicspath{{Figs/}}

\shorttitle{Magnetic Analyses of Active Region 11117}
\shortauthors{Jiang et al.}

\begin{document}
\begin{CJK*}{UTF8}{gbsn}


  \title{Study of the 3D Coronal Magnetic Field of Active Region 11117
    Around the Time of a Confined Flare Using a Data-Driven CESE--MHD
    Model}


\author{
  Chaowei Jiang (江朝伟)\altaffilmark{1}, 
  Xueshang Feng (冯学尚)\altaffilmark{1},
  S.~T. Wu (吴式灿)\altaffilmark{2}, 
  Qiang Hu (胡强)\altaffilmark{2}} 
\email{cwjiang@spaceweather.ac.cn}


\altaffiltext{1}{SIGMA Weather Group, State Key Laboratory for Space
  Weather, Center for Space Science and Applied Research, Chinese
  Academy of Sciences, Beijing 100190} 

\altaffiltext{2}{Center for Space Plasma and Aeronomic Research, The
  University of Alabama in Huntsville, Huntsville, AL 35899, USA}


\begin{abstract}
  We apply a data-driven MHD model to investigate the
  three-dimensional (3D) magnetic field of NOAA active region (AR)
  11117 around the time of a C-class confined flare occurred on 2010
  October 25. The MHD model, based on the spacetime
  conservation-element and solution-element (CESE) scheme, is designed
  to focus on the magnetic-field evolution and to consider a
  simplified solar atomsphere with finite plasma $\beta$. Magnetic
  vector-field data derived from the observations at the photoshpere
  is inputted directly to constrain the model. Assuming that the
  dynamic evolution of the coronal magnetic field can be approximated
  by successive equilibria, we solve a time sequence of MHD equilibria
  basing on a set of vector magnetograms for AR 11117 taken by the
  Helioseismic and Magnetic Imager (HMI) on board the {\it Solar
    Dynamic Observatory (SDO)} around the time of flare. The model
  qualitatively reproduces the basic structures of the 3D magnetic
  field, as supported by the visual similarity between the field lines
  and the coronal loops observed by the Atmospheric Imaging Assembly
  (AIA), which shows that the coronal field can indeed be well
  characterized by the MHD equilibrium in most time. The magnetic
  configuration changes very limited during the studied time interval
  of two hours. A topological analysis reveals that the small flare is
  correlated with a bald patch (BP, where the magnetic field is
  tangent to the photoshpere), suggesting that the energy release of
  the flare can be understood by magnetic reconnection associated with
  the BP separatrices. The total magnetic flux and energy keep
  increasing slightly in spite of the flare, while the computed
  magnetic free energy drops during the flare with an amount of $\sim
  10^{30}$~erg, which seems to be adequate to provide the energy
  budget of the minor C-class confined flare.
\end{abstract}


\keywords{Magnetic fields; Magnetohydrodynamics (MHD); Methods:
  numerical; Sun: corona}



\section{Introduction}
\label{sec:intro}

The magnetic field holds a central position within solar research such
as sunspots, coronal loops, prominences, and spectacular solar
phenomena like flares and coronal mass ejections (CMEs). It has been
commonly accepted that the energy released by solar flare (which is
usually up to the order of $10^{32}$~erg during the major events) must
be sourced from magnetic field of the active region since all the
other possible energy sources are completely inadequate
\citep{Priest1987Book}. To help quantitative understanding the solar
explosive phenomena such as flare and CMEs, it is essential to get the
knowledge of the amount of free magnetic energy and its temporal
variation during the events. Magnetic reconnection is attributed by
most flare models to the basic mechanism for energy conversion rapidly
from the magnetic field into the kinetic and thermal counterparts
\citep{Priest2002,Shibata2011}. To locate where magnetic reconnection
is prone to happen and produce flare needs the three-dimensional (3D)
coronal magnetic field, by which the important topological and
geometrical features that are favorable sites for reconnection, e.g.,
the null point, the separatrices or more commonly the quasi-separatrix
layers \citep{Priest2002,Titov2002,Longcope2005}, can then be
found. Unfortunately, the 3D magnetic field in the corona is very
difficult to be directly observed, although information of the 3D
geometrical configuration of the field lines can be partially
reconstructed by the method of stereoscopy using coronal loops
observed in different aspect angles in the EUV and X-ray wavelengths
\citep[see living review by][]{Aschwanden2011}. Up to the present, a
routinely direct measurement of the solar magnetic field on which we
can rely is mainly restricted to the solar surface, i.e., the
photosphere \citep[there are only a few cases available with the
measurements of the chromospheric and coronal fields, e.g.,][]
{Solanki2003,Lin2004}.

With in hand the observed magnetic field on the photosphere, there are
several ways to study the evolution of 3D magnetic field in the
corona. One of them is the well-known model of field extrapolation
from the magnetogram, especially, the nonlinear force-free (NLFF)
field extrapolation
\citep{Wiegelmann2008,Schrijver2008,Derosa2009}. As the solar corona
is dominated by the magnetic field environment with very small plasma
$\beta$ (the ratio of gas pressure to magnetic pressure), the force
free model is usually valid and serves as a good approximation for the
low corona (but above the photosphere) in a near-static state. A
variety of numerical codes have been developed to implement the
force-free field extrapolation in the past decade
\citep[e.g.,][]{Wheatland2000,Wiegelmann2004,Amari2006,Jiang2012apj}. These
methods have been applied to analyze the magnetic structures of the
active regions, the electric current distributions, energy budget of
the eruptions, etc
\citep{Regnier2006,Guo2008,Thalmann2008,Jing2010,Valori2012,Sun2012},
with success made, such as reproducing the field lines comparable with
the observed coronal loops \citep[e.g.,][]{Wiegelmann2012} and
extrapolating complex flux rope which is believed to be associated
with the filament channel \citep[e.g.,][]{Canou2010}. However, it
should be noted that the success is still limited when applied to
realistic solar data \citep{Schrijver2008,Derosa2009,Schrijver2009},
which is mainly because of the intrinsic non-force-freeness in the
field close to the photosphere \citep{Metcalf1995}. Thus the observed
data can generally not provide a consistent boundary condition for the
model based on an exact force-free assumption, and some {\it ad hoc}
preprocessing (to {\it remove} the force in the raw magnetogram) is
usually made to prepare the vector magnetograms for the extrapolation
codes \citep{Wiegelmann2006b}.

Another method is using data-driven magnetohydrodynamics (MHD) model
which is more general than the force-free one
\citep{Wu2006,Wu2009,Wang2008ApJ,Jiang2010,Fan2011,Fan2012RAA}. This
is because in the MHD model, nonlinear dynamic interactions of the
magnetic field and plasma flow field are treated in a self-consistent
way, in which the near force-free state of the coronal magnetic field
is included. A first data-driven MHD model was developed by
\citet{Wu2006} for simulating the evolution of active regions. In
their original work, the initial setup of the model is established by
seeking a MHD equilibrium started from an arbitrarily prescribed
plasma and a potential magnetic field based on the {\it Solar and
  Heliospheric Observatory (SOHO)}/MDI magnetogram at a given
time. Then a time-series of MDI magnetograms observed afterward were
continuously inputted at the bottom boundary to drive the above field
to respond to the changes on the photosphere. In particular, the
procedure of continuously feeding observed data on the bottom boundary
is made to be self-consistent by a projected-characteristic method
\citep[e.g.,][]{Nakagawa1981,Wu1987}.

If in an ideal or strict condition, this dynamic process of the
data-driven model can indeed be regarded as evolution of the
corona. However, in the reality, there are still many difficulties and
problems in using a data-driven MHD model to study the active region
evolution. First of all is the lack of observations for the
photospheric parameters of plasma such as the surface flow velocity,
which is important boundary information for the driving process
\citep{Abbett2004,Welsch2004}. This is especially essential by
regarding that at the photosphere the magnetic field may be dominated
by the dense plasma (with high $\beta$), and the field lines anchored
in the photosphere can usually be considered as line-tied by the
photospheric plasma because of the high electric conductivity
\citep{Priest1987Book,Mikic1994,Solanki2006}. This means that the
field-line footpoints are passively advected by the plasma flow which
itself is induced in the convection zone below. Without the
information of the surface flow, response of the coronal field lines
driven by photospheric footpoint-motion cannot be fully followed. This
encourages people to recover the photospheric flow velocity from the
time-varying magnetograms by using local correlation tracking
technique or similar methods \citep[e.g.,
see][]{Chae2001,Welsch2004,Demoulin2009}. The second problem comes
from the cadence of the observed data which is generally too low for a
data-driven model that needs a highly continuous data flow. To address
this problem, \citet{Wu2006} simply used a time-linear interpolation
on the 96-minute cadence MDI magnetograms to provide the data needed
at each time step \citep[about 6-second used by][]{Wu2006} of the
model. This obviously over-simplifies the real evolution of the
photospheric field which is very time-nonlinear, but it may be the
only choice one can made \footnote{This problem can be alleviated now
  by using the recently available data recorded by HMI on-board the
  new observatory {\it SDO}, which has a higher data cadence.}. In
view of these two problems, it may be more practical to construct
independent MHD equilibrium for each one of the magnetograms and
consider these successive equilibria as the continuous time-evolution
of the corona, as done by \citet{Wu2009,Fan2011}.

For the third problem, it is difficult to couple the photospheric and
the coronal plasma in a single model because of the highly stratified
plasma, of which the parameters, i.e., the density and temperature
change drastically by several orders of magnitude within an extremely
thin layer (the chromosphere and transition region) above the
photosphere due to some kind of unknown coronal heating process. As a
realistic model with inputted magnetic field data observed on the
photosphere, it is required to describe the behavior of the magnetic
field in this stratified environment with plasma $\beta$ varying from
$> 1$ (the photosphere) to $\ll 1$ (the corona). However, this
challenges greatly the numerical scheme and computational resource to
treat the transition region and additionally, one may need to
incorporate the complicated thermodynamic processes of the real
corona, such as the thermal conduction and radiative losses
\citep[e.g., see models by][]{Abbett2007,Fang2010}. We note that in
the works of \citep{Wu2006,Wu2009,Wang2008ApJ,Fan2011}, only the
photospheric or near-photosphere plasma is considered in the models
and thus these models are mainly used for studying the evolution of
photospheric parameters, such as the plasma flow, the Poynting flux,
the current helicity and some other non-potential parameters at the
photosphere level. The evolution of the 3D coronal magnetic field, on
the other hand, was rarely studied by using these models because of
the unjustified high-$\beta$ and dense plasma environment. This is due
to the reason mentioned above that a coupled modeling of the
photospheric and coronal fields is still computationally prohibitive.

In this work we will use the data-driven MHD model to study the 3D
coronal field within a low plasma--$\beta$ condition. The numerical
model is developed following our previous work \citep{Jiang2011},
which has been devoted to a validation of the CESE--MHD method for
reconstructing the 3D coronal fields using a semi-analytic force-free
field solution proposed by \citet{Low1990}. We will study the 3D
magnetic field and its evolution of active region NOAA AR 11117 around
the time of a small C-class flare happened on 2010 October 25,
observed by {\it SDO}/AIA with a time-series of vector-magnetograms
recorded by {\it SDO}/HMI. While the 3D magnetic field of the same
active region has been studied by \citet{Sun2010} and
\citet{Tadesse201211117} using the NLFFF model, this is the first
study we apply the CESE--MHD model to realistic solar data. Similarly
assuming that the evolution of the coronal magnetic field in active
region can be described by successive equilibria
\citep[e.g.,][]{Regnier2006,Wu2009,Sun2012,Tadesse201211117}, we use
each vector-magnetogram of the data set to get a snapshot MHD
equilibrium and study the temporal evolution of the field by a series
of these equilibria. This method is justified by considering that the
evolution of the active region, driven by the photospheric motion with
flow speed on the order of several km~s$^{-1}$, is sufficiently slow
compared with the speed of the coronal magnetic field relaxing to
equilibrium, which is up to thousands of km~s$^{-1}$
\citep{Antiochos1987,Seehafer1994}. It is also valid for the present
studied objective, the AR 11117, which shows no major changes of the
magnetic field in the chosen time period.

The remainder of the paper is organized as follows. In
Section~\ref{sec:model} we give a brief description of the CESE--MHD
model. Magnetic field data used for driven the model is described in
Section~\ref{sec:data}. The modeling result for the AR 11117 is
presented in Section~\ref{sec:res}, including a qualitative inspection
of the the 3D magnetic configurations, topological analysis of the
field at the flare site, as well as study of the magnetic energy
budget and current distribution. Finally we draw conclusions and give
some outlooks for future work in Section~\ref{sec:conclude}.

\section{The Data-Driven CESE--MHD Model}
\label{sec:model}

In a nutshell, what we intend to solve is a set of MHD equilibria of
which each is consistent with a snapshot of magnetic field observed on
the photosphere. We thus start from an arbitrarily initial field,
e.g., a potential or linear force-free field, with a plasma and input
at the bottom of the model the vector magnetogram to drive the system
away from its initial state and then let the system relax to a new
equilibrium. The numerical model follows our previous work
\citep{Jiang2011}. Since the computation is focused on the magnetic
field and its dynamics with plasma in the low corona, here we use a
simplified solar atmosphere with a low plasma $\beta$ and a uniform
constant temperature. Thus the numerical scheme need only to handle
the plasma density $\rho$, the flow velocity $\vec v$ and the magnetic
field $\vec B$. The MHD equations are written as follows:
\begin{eqnarray}
  \label{eq:main_equ}
  \frac{\partial \rho}{\partial t}+\dive (\rho\vec v) = 0,
  \nonumber \\
  \rho\frac{D\mathbf{v}}{D t} = -\grad p+\vec J\times \vec B+\rho\vec
  g + \nabla\cdot(\nu\rho\nabla\mathbf{v})-\nu_{f}\rho\vec v,
  \nonumber \\
  \frac{\partial\mathbf{B}}{\partial t} = 
  \nabla\times(\mathbf{v}\times\mathbf{B}).
\end{eqnarray}
In these equations: $\vec J$ is the electric current; $p$ is the gas
pressure given by $p=\rho R T_{0}$ where $R$ is gas constant and
$T_{0}$ is the constant temperature; $\vec g$ is the solar gravity and
is assumed to be constant as its photospheric value since we simulate
the low corona with height of about $100$~Mm from the photosphere. A
small kinematic viscosity $\nu$ with a value of $\sim \Delta
x^{2}/\Delta t$ ($\Delta x$, $\Delta t$ are respectively the grid
spacing and the time step in the numerical scheme) is added for
consideration of numerical stability.

Different from the equations used in \citet{Jiang2011}, here we
include an additional frictional force $-\nu_{f}\rho\vec v$ to deal
with the problem that in some odd places near the magnetogram (i.e.,
the bottom) the plasma velocity is prone to be accelerated to
extremely high due to very large gradients or some kind of
uncertainties intrinsically contained in the observed data. This is
because the data is very intermittent in the observed magnetograms,
which usually show a large number of small-scale polarities and even
apparent discontinuities, and these features cannot be adequately
resolved by the grid resolution. We find that such problems can
severely restrict the time step and slow the relaxation process of the
entire system, even making the computation unmanageable. It should be
noted that including the friction force is only an {\it ad hoc} choice
for numerical consideration in the case of dealing with the original
data in the model. Alternatively, one can perform certain smoothing on
the original magnetograms beforehand to remove noise and decrease
large gradients in the raw data. This, however, may erase some of the
important parasitic polarities around the major sunspots and also
probably change the locations of the polarity inversion lines (PILs),
which could influence the analysis of the local field configurations
responsible for small-scale energy dissipation near the photosphere
(e.g., the small flare in the present study). Also there is magnetic
flux loss and the energy content of the field may be affected, if the
vertical component of the magnetogram is modified
\citep{Metcalf2008}. Although the field at the coronal base ought to
be smoother than the photospheric field because of the field expansion
from the high-$\beta$ to the low-$\beta$ regions, in which way such
smoothness can be modeled is still problematic. To this end, it is
prudent not to smooth the original magnetograms and thus we use the
frictional force to control the above-mentioned problem in the
numerical computation. We have tried different values for the
frictional coefficient $\nu_{f}$ and adopt a optimized one $\nu_{f} =
1/(50\Delta t)$, which can control the plasma flow in a reasonable
level, i.e., the flow speed is suppressed under the maximum Alfv\'en
speed but not too small. Our tests show that the adjustment of
$\nu_{f}$ affects the MHD relaxation process but gives almost the same
final solution. Finally, no explicit resistivity is included in the
magnetic induction equation, since the numerical diffusion can lead to
topological changes of the field when necessary.

\begin{figure*}[htbp]
  \centering
  \includegraphics[width=\textwidth]{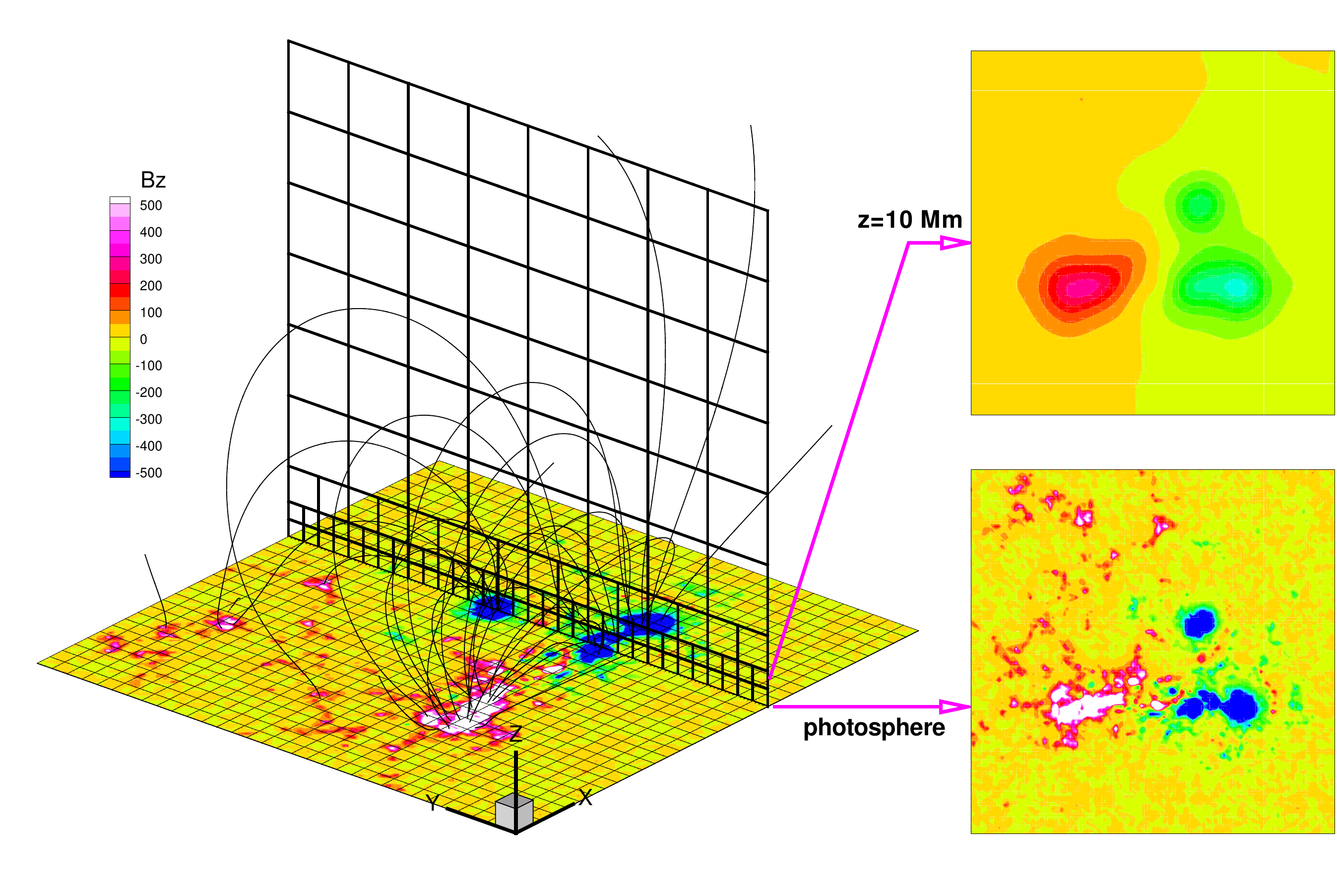}
  \caption{The configuration of the computational grid. The entire
    volume is divided into blocks and each block has $8\times 8\times
    8$ cells. In the left panel, two slices through the volume are
    plotted to show the structure of the blocks, which are outlined by
    the black grid lines; the bottom contour map represents $B_{z}$ on
    the photosphere and the curved lines show the potential field
    lines. The right panels show the 2D contour images of $B_{z}$
    sliced at $z=0$ and $z=10$~Mm (locations in the 3D grid are shown
    by the arrows).}
  \label{fig:grid}
\end{figure*}

The above equation system~(\ref{eq:main_equ}) is solved by our
CESE--MHD code \citep{Jiang2010}. The CESE method deals with the 3D
governing equations in a substantially different way that is unlike
traditional numerical methods (e.g., the finite-difference or
finite-volume schemes). The key principle, also a conceptual leap of
the CESE method, is treating space and time as one entity. By
introducing the conservation-element (CE) and the solution-element
(SE) as the vehicles for calculating the spacetime flux, the CESE
method can enforce conservation laws both locally and globally in
their natural spacetime unity form. Compared with many other numerical
schemes, the CESE method can achieve higher accuracy with the same
mesh resolution and provide simple mathematics and coding free of any
type of Riemann solver or eigendecomposition. For more detailed
descriptions of the CESE method for MHD simulations including the
multi-method control of the well-known $\divB$ numerical errors,
please refer to our previous work, e.g., \citet{Feng2006,Feng2007},
\citet{Jiang2010} and \citet{Jiang2011}. We use a non-uniform grid
within the framework of a block-structured, distributed-memory
parallel computation. The grid configuration is depicted in
\Fig~\ref{fig:grid}. Specifically, the whole computational volume is
divided into blocks with different spatial resolution, and the blocks
are evenly distributed among the CPU processors. In the $x$--$y$
plane, i.e., the plane parallel with the photosphere, the blocks have
the same resolution. In the vertical direction, resolutions of the
blocks are decreased with height, e.g., near the photosphere, the grid
spacing matches the resolution of the magnetogram and up to the top of
the model box, the grid spacing is increased by four times. As shown
by \Fig~\ref{fig:grid}, at a height of only 10~Mm the magnetic field
has become far less intermittent, i.e., much smoother than that at the
photosphere. Thus using this non-uniform mesh can affect the
computational accuracy little compared with a uniform mesh, but can
save significant computational resources.

\begin{figure}[htbp]
  \centering
  \includegraphics[width=0.48\textwidth]{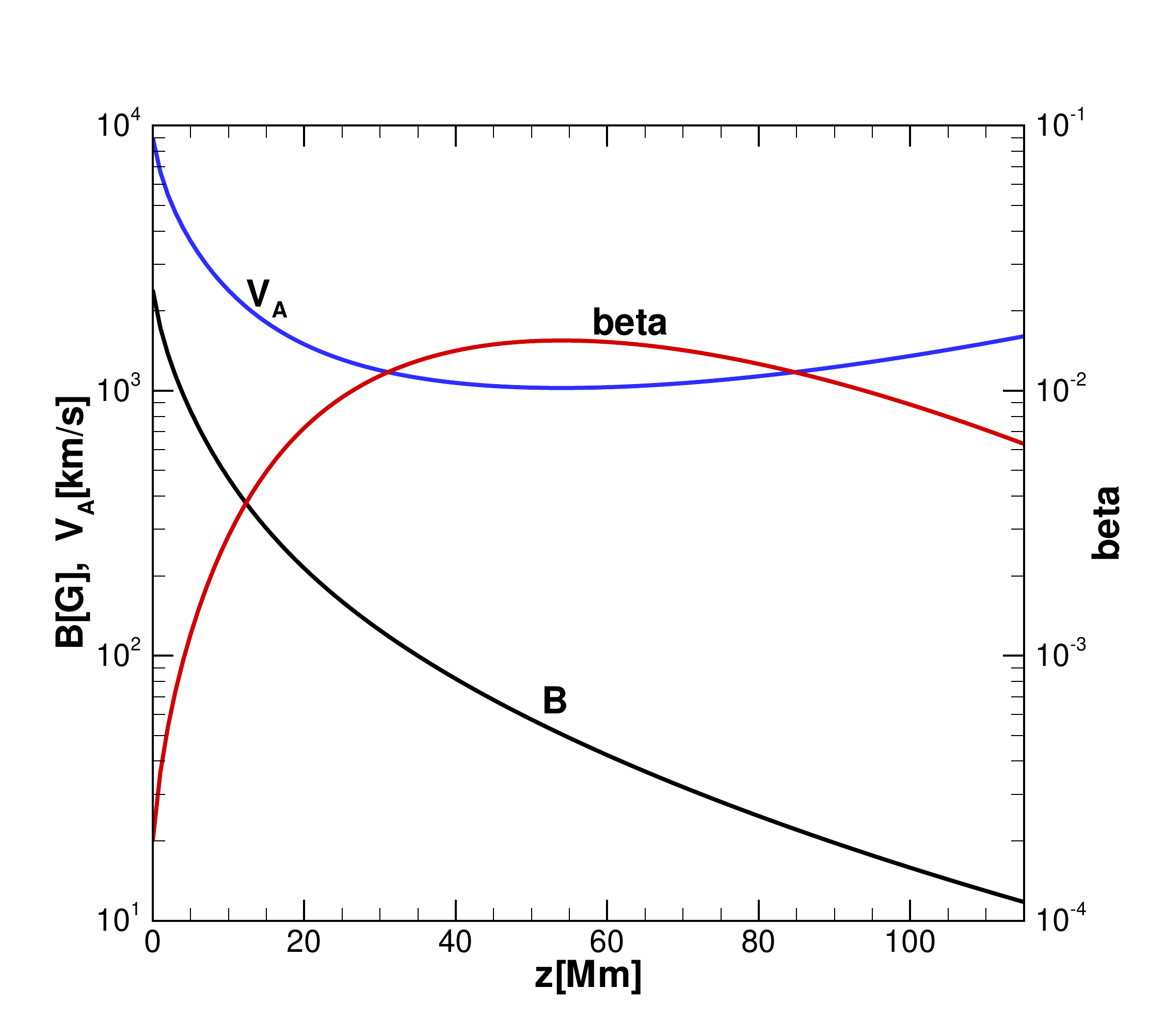}
  \caption{Typical configurations of the magnetic field strength $B$,
    the Alfv\'en speed $V_{\rm A}$ and the plasma $\beta$ along a
    vertical line through the computation volume.}
  \label{fig:params}
\end{figure}

The initial configuration of the simulation consists of a potential
field matching the vertical component of the magnetogram and a plasma
in hydrostatic equilibrium in the solar gravitational field. The
potential field is obtained by a Green's function method
\citep[e.g.,][]{Metcalf2008}. The plasma density is given by $\rho(z)
= \rho_{0}\exp(-z/H)$ where $H$ is the pressure scale height $H =
RT_{0}/g$ and $z=0$ denotes the photosphere. Nondimensionalization of
the parameters is same as \citet{Jiang2011} and \Fig~\ref{fig:params}
shows a typical configuration of the parameters along a vertical line
through the computation volume at the strong magnetic region. It is
noteworthy that the plasma $\beta$ can be large in the relatively weak
field region and thus the intrinsic force in the vector magnetogram
can be self-consistently balanced by the plasma in the MHD relaxation
process. This is unlike the force-free model, as aforementioned (see
Section~\ref{sec:intro}), which generally cannot deal with the
observed data directly. The boundary conditions are also very similar
to those used in \citet{Jiang2011}: the bottom boundary is fed with
the observed vector magnetogram incrementally in tens of Alfv\'en time
until the observed data is fully matched, and all other boundaries are
set by the non-reflecting boundary conditions. Besides, the flow
velocity on the bottom is set by extrapolation from the neighboring
inner grid. This has a function to increase the communication between
the magnetogram and the computational volume \citep{Valori2007}.

\section{Data}
\label{sec:data}

\begin{figure*}[htbp]
  \centering
  \includegraphics[width=\textwidth]{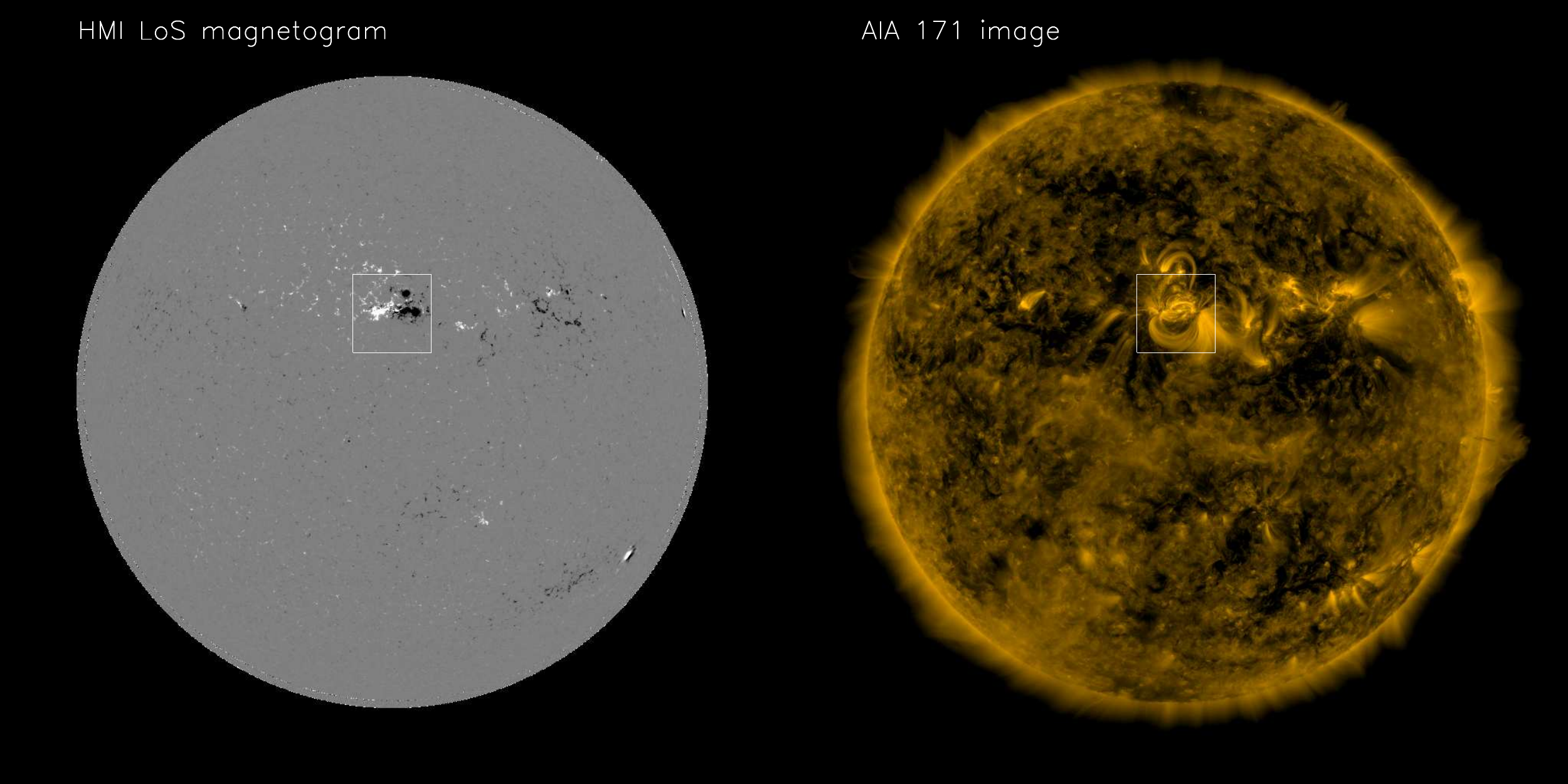}
  \caption{Full-disk {\it SDO}/HMI line-of-sight (LoS) magnetogram
    (left) and full-disk {\it SDO}/AIA 171 {\AA} image. Both images
    are obtained at the same time of 22:12 UT on 2010 October 25, and
    have been co-aligned. AR 11117 is outlined by the white rectangle
    on the images.}
  \label{fig:hmi_fulldisk}
\end{figure*}

\begin{figure*}[htbp]
  \centering
  \includegraphics[width=\textwidth]{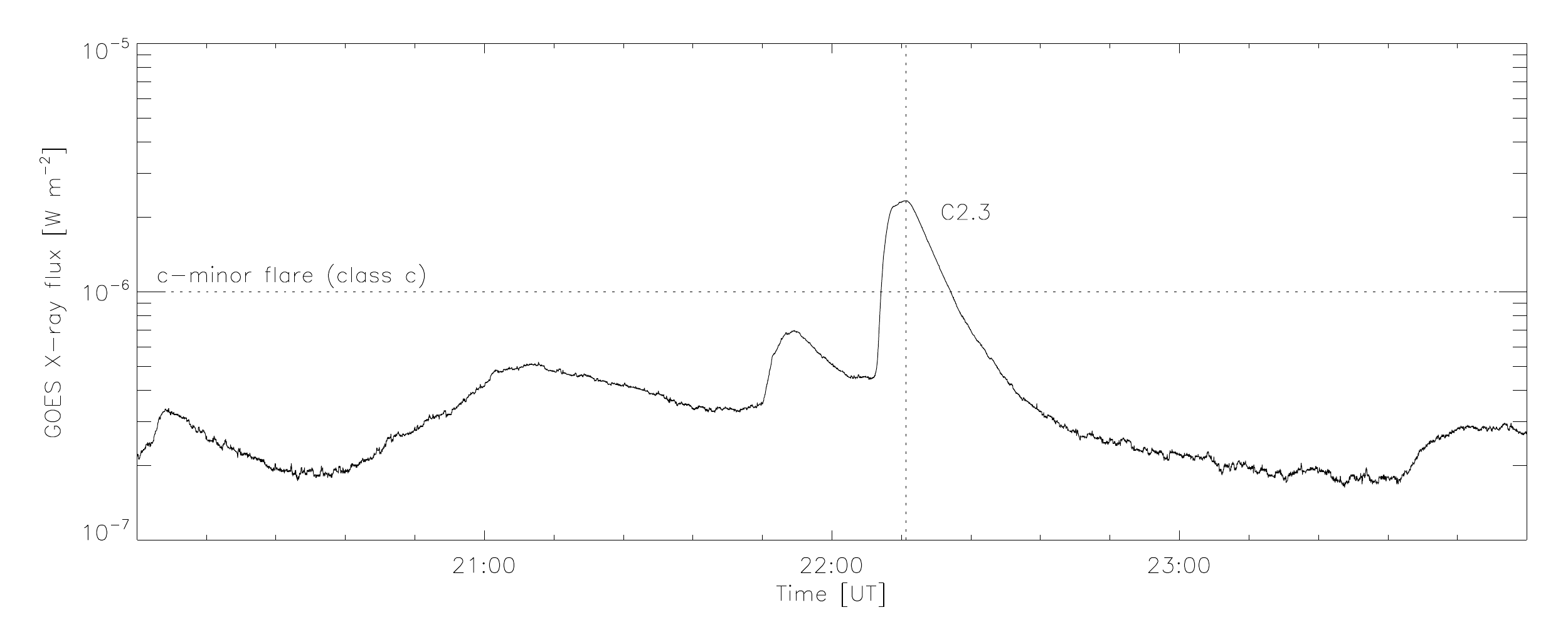}
  \caption{GOES soft-X ray flux from 20:00 UT to 24:00 UT on 2010
    October 25 in the wavelength range of 1--8~{\AA}. Horizontal
    dotted line indicates the C-minor flare class and the vertical
    dotted line indicates the peak time of the flux.}
  \label{fig:goes}
\end{figure*}

Active region NOAA AR 11117 was observed by SDO from 2010 October 20
to 2010 November 2, mainly during Carrington Rotation 2102. On 2010
October 25, it was crossing the central meridian of the solar disk
with latitude of $22^{\circ}$ as shown in the full-disk HMI and AIA
images (\Fig~\ref{fig:hmi_fulldisk}). On this date solar activity
was dominated by AR 11117 with many small B-class flares observed and
near the end of the day, a C2.3-class flare happened. NOAA records
indicate that the event began in soft X-rays (SXRs) which were
detected by the GOES (Geostationary Operational Environmental
Satellite) 15 satellite at 22:06 UT, reaching a peak at 22:12 UT and
ending at 22:18 UT (see \Fig~\ref{fig:goes}). As observed by AIA (see
\Fig~\ref{fig:AIA_compare}), the central part of the active region
shows distinct brightenings at the flare peak time and the flare is
confined in rather low altitude without inducing major changes in the
coronal loops or eruptions.


\begin{figure*}[htbp]
  \centering
  \includegraphics[width=\textwidth]{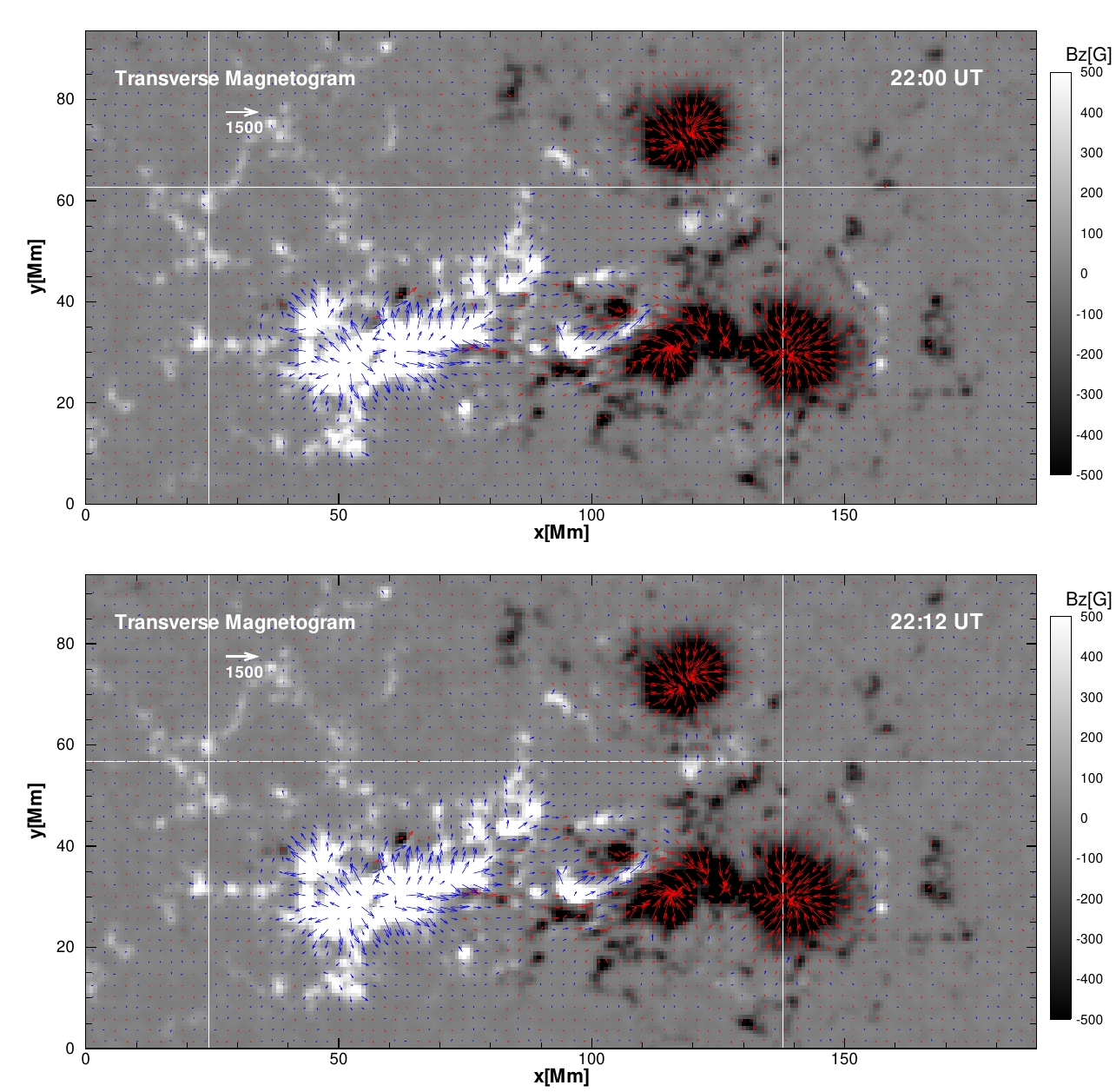}
  \caption{Vector magnetograms for AR 11117 at time of 22:00 and
    22:12. The gray images represent $B_{z}$ with a saturation level
    of $\pm 500$~G. The tangential fields are shown by the vectors
    (plotted at every third pixel point) with blue color in the
    positive $B_{z}$ region and red in the negative $B_{z}$ region.}
  \label{fig:magnetogram}
\end{figure*}

We select a set of vector magnetograms for AR 11117 which were taken
by HMI around the flare peak time with a cadence of roughly half an
hour.  The data is de-rotated to the disk center, the field vectors
are transformed to Heliographic coordinates with projection effect
removed and finally remapped to a local Cartesian coordinate using
Lambert equal area projection. For a detailed processing of the HMI
vector magnetograms please refer to
\url{http://jsoc.stanford.edu/jsocwiki/VectorMagneticField}.
Specifically, six magnetograms taken at 21:00, 21:36, 22:00, 22:12,
22:36, and 23:00 UT, respectively, are used for the
simulation. \Fig~\ref{fig:magnetogram} shows examples of the vector
magnetograms before and at the flare peak time (the gray image shows
the vertical component $B_{z}$ and the arrows indicate the transverse
field). There are four regions with flux greater than 1000~G
concentrated in areas of about 10 arcsec square and these regions are
manifested as four main sunspots observed in the AIA 4500 {\AA} (white
light) image. Strong shear of the transverse field can be seen near
the image center, with the vector almost parallel to the PIL (see the
regions where the color of the vectors changes while their directions
are nearly the same). The original resolution of the magnetogram is
about 0.5~arcsec ($\sim 360$~km) per pixel and we bin the data to
1~arcsec~pix$^{-1}$ for our simulation with a field of view of
$256\times 256$~arcsec$^{2}$ ($184\times 184$~Mm$^{2}$). The height of
the computational box is set to 160~arcsec ($115$~Mm). To reduce the
side and top boundary influence, the following analysis of the results
is performed on a subvolume with $200\times 128\times
100$~arcsec$^{3}$ centered in the full computational domain.

\section{Results}
\label{sec:res}

\subsection{Comparison with AIA Loops}

\begin{figure*}[htbp]
  \centering
  \includegraphics[width=\textwidth]{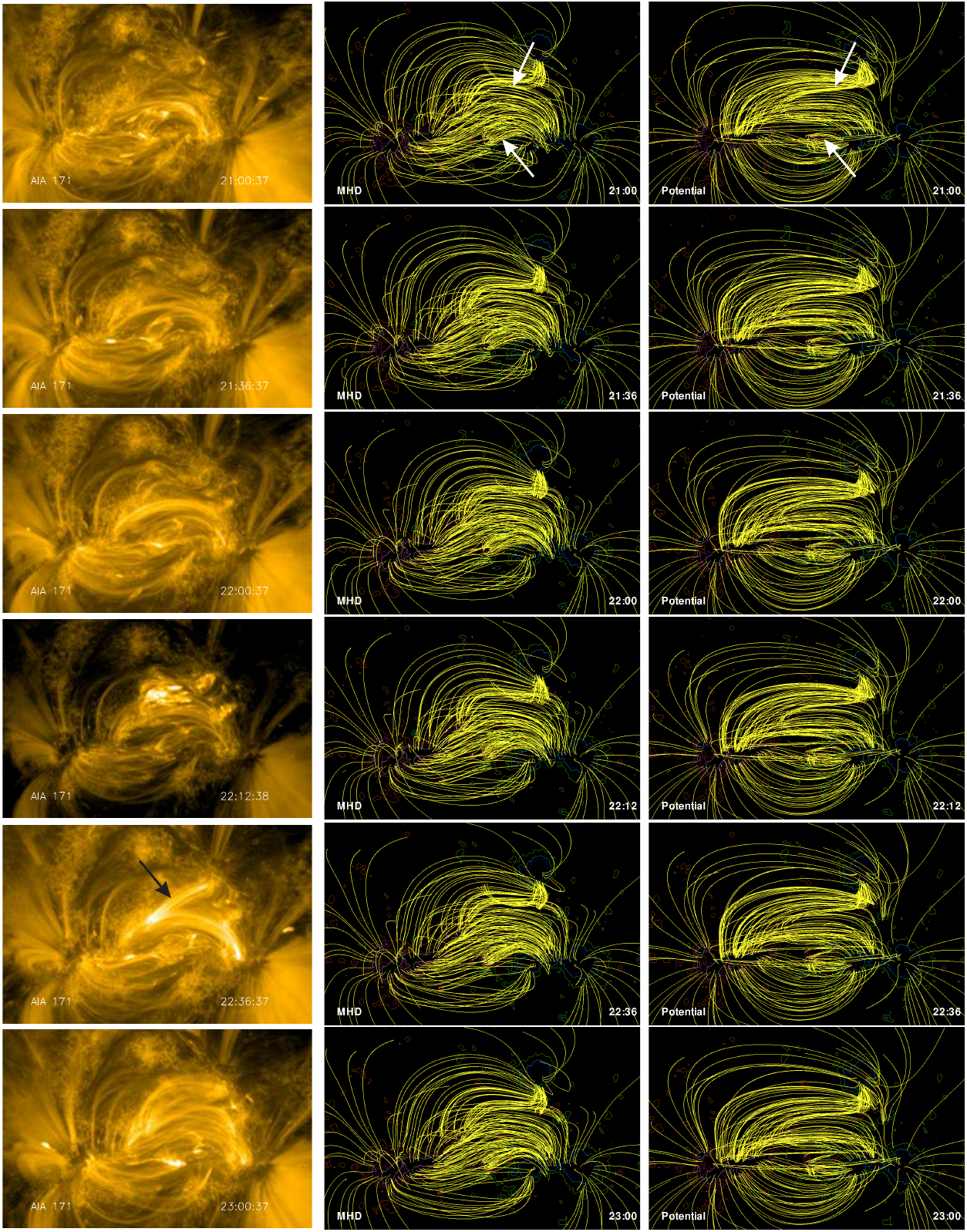}
  \caption{Comparison of the modeled field lines with {\it SDO}/AIA
    171~{\AA} image for AR 11117. The left column is the AIA images,
    the middle column is the selected magnetic field lines from the
    MHD results. Also the potential field lines are plotted in the
    last column. The field lines in all the panels are traced from the
    same footpoints on the photosphere and the color contours of
    photospheric $B_{z}$ are plotted on the background.}
  \label{fig:AIA_compare}
\end{figure*}

The high-resolution coronal loops observed by {\it SDO}/AIA in the
wavelength of 171~{\AA} give us a proxy of the magnetic field-line
geometry (see the left column of \Fig~\ref{fig:AIA_compare}) and also
a good constraint for the magnetic field model. In the middle column
of \Fig~\ref{fig:AIA_compare} we show some selected magnetic field
lines of the model results. In these images, the yellow lines
represent the magnetic field lines and the background contours outline
the vertical component of the magnetogram. For a visual comparison
with the observed coronal loops, we plot the figures side-by-side with
the AIA 171~{\AA} images observed at the same time. The field lines
are selected roughly according to the visible bright loops and the
angle of view of the MHD results is co-aligned with the AIA image. As
shown from an overview of the figures, the simulated field lines
resemble quite well the observed loops, especially at the central
region of the AR where the field lines are sheared strongly. This
means that the field there is very non-potential. The potential fields
at each time are shown in the third column of
\Fig~\ref{fig:AIA_compare}. Compared with the potential field lines,
the MHD field lines exhibit some twists, although not strong, implying
the existence of field-line-aligned currents (i.e., currents along the
field lines). We find that there are features well reproduced by the
MHD model but failed to be recovered by the potential model, for
example, the structures pointed out by the white arrows in the
figure. Reconstruction of these features, obviously needing variation
of the field-line connectivities from those of the initial potential
field, demonstrates that our model can indeed reconstruct the 3D
magnetic topology which is implied in the observed transverse field.

By comparison we conclude that within this AR, the MHD model gives
much better results than the potential field model. Although during
this time interval some small changes can be recognized in the loops
and the MHD field lines, it is difficult to find any variation in the
magnetic topology around the time of flare from 22:00 to 22:12. This
means that the flare-related reconnection takes place at rather small
scale and low height near the photosphere (as indicated by the
analysis of magnetic topology in the following section). It can be
clearly seen in the AIA image at 22:36 there are two groups of loops
with much more brightness than the other loops, and the MHD model
appears to fail to reproduce the groups in the north (i.e., the loops
pointed out by the black arrow in the image, note that this group of
loops is difficult to find in all the other AIA images in the
figure). This may be due to that the central part of the active region
is very dynamic after the flare since hot plasma from the chromosphere
`evaporated' to the post-flare loops, thus cannot be well described by
the quasi-static state which we have sought.

\subsection{Topological Analysis of the Flare Location}

It has been thought that flare is plausible to take place in the
regions with strong variation of the field line connectivity
\citep[e.g.,][]{Mandrini1995,Demoulin1997}. Such regions are called
quasi-separatrix layers (QSLs), which are generalized from the concept
of magnetic separatrices where the field-line linkage (or
connectivity) is discontinuous \citep{Priest1995,Demoulin1996}. To
locate the QSLs in the 3D coronal field, \citet{Titov2002} introduced
a so-called {\it squashing factor} ($Q$) to quantify the change of the
field linkage basing on the field-line mapping. For the corona field,
the mapping is defined from one photospheric footpoint $(x,y)$ of a
given field line to the other photospheric footpoint
$(X(x,y),Y(x,y))$, which is also called the magnetically conjugate
footpoint\footnote{Here we need not to distinguish the footpoints of
  the positive and negative polarities, since $Q$ is designed with the
  same value at conjugate footpoints of the same field line
  \citep{Titov2002}.}. Then the squashing factor is given by
\begin{equation}
  \label{eq:Q}
  Q = \frac{a^{2}+b^{2}+c^{2}+d^{2}}{|ad-bc|}
\end{equation}
where 
\begin{equation}
  a = \frac{\partial X}{\partial x},\ \ 
  b = \frac{\partial X}{\partial y},\ \ 
  c = \frac{\partial Y}{\partial x},\ \ 
  d = \frac{\partial Y}{\partial y}.
\end{equation}
Producing a map of $Q$ factor is a robust way to find the topological
elements (including both the QSLs and the separatrices) in the 3D
magnetic field, but its calculation is a computational intensive work
because field lines are required to be traced from not only each point
but also its neighboring points to estimate the derivatives of the
field line mapping. We thus use the following algorithm: first, field
line from each grid point $(i,j)$ on the bottom surface is traced
either forward or backward and location of the other (conjugate)
footpoint is denoted by $(X(i,j),Y(i,j))$; then at each grid point a
centered difference involving with its neighboring four grid points
$(i-1,j),(i+1,j),(i,j-1),(i,j+1)$ is used to approximate the elements
needed by $Q$, i.e.,
\begin{eqnarray}
  \label{eq:abcd}
  a = \frac{X(i+1,j)-X(i-1,j)}{2\Delta x},\nonumber\\ 
  b = \frac{X(i,j+1)-X(i,j-1)}{2\Delta y},\nonumber\\ 
  c = \frac{Y(i+1,j)-Y(i-1,j)}{2\Delta x},\nonumber\\
  d = \frac{Y(i,j+1)-Y(i,j-1)}{2\Delta y}.  
\end{eqnarray}
where $\Delta x$ and $\Delta y$ are the grid spacings. To avoid the
numerical uncertainties of tracing field lines with very small
structures near the photosphere (i.e., structures smaller than the
grid resolution), we raise the bottom surface by three pixels (about
2~Mm) above the photosphere in the computation. This might smooth out
some very fine structures in the $Q$ map but the basic topological
features remain since they depend mainly on the large-scale current
distribution \citep{Titov1999}. As suggested by \citet{Titov2002}, it
is also useful to compute the expansion-contraction degree ($K$) which
is defined by the ratio of the normal components of the magnetic field
at the two ends of the field lines. While the factor $K$ has similar
function to locate the QSLs as $Q$, it is much simpler to compute than
the latter, and may be more reliable since its computation is free of
the numerical errors of the finite difference in \Eq~(\ref{eq:abcd}).

\begin{figure*}[htbp]
  \centering
  \includegraphics[width=\textwidth]{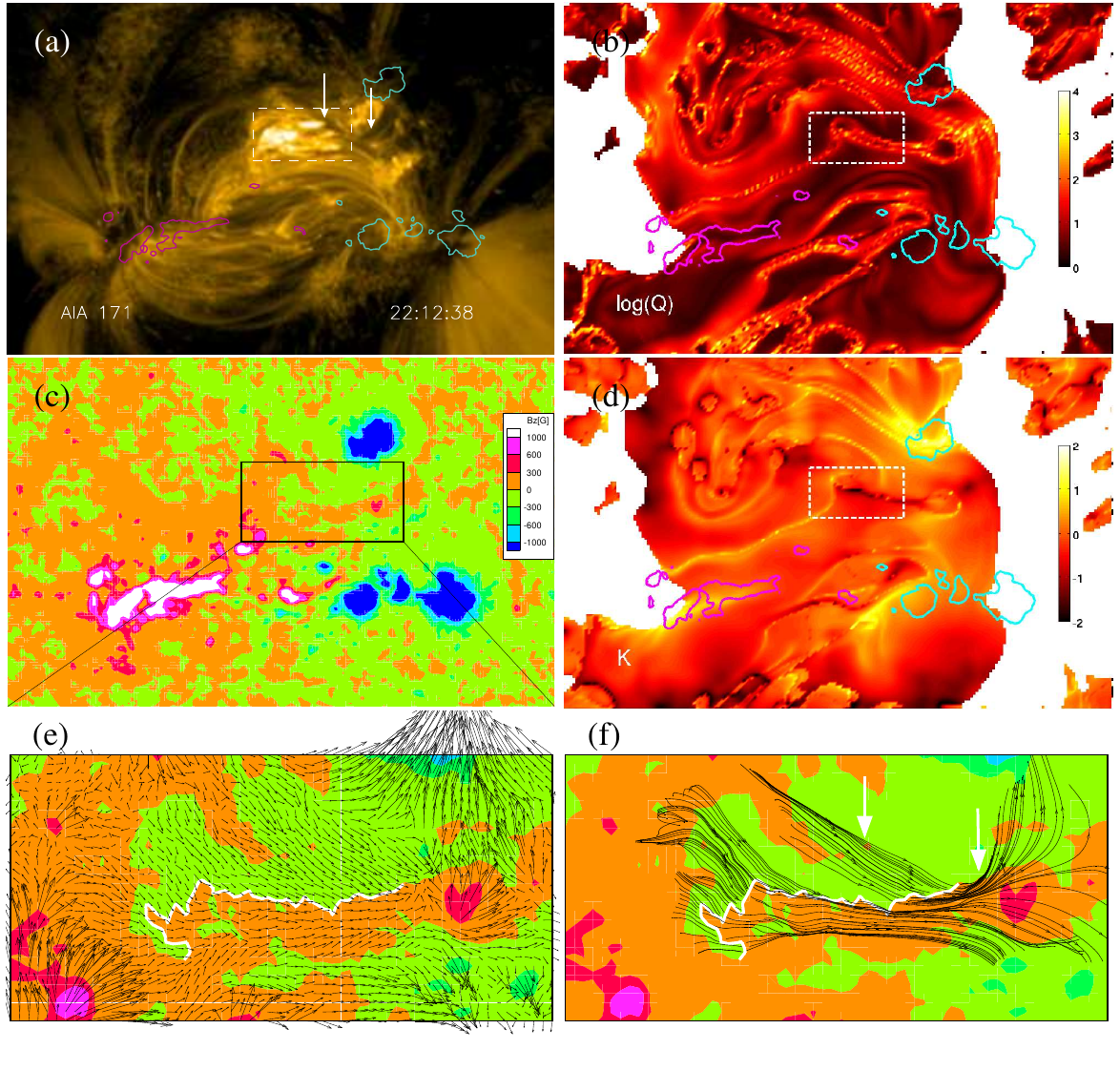}
  \caption{(a) The AIA 171 image at 22:12, with the contour lines
    plotted for the LoS photospheric field at $\pm 1000$~G and the
    dashed box showing the location of flare with loops
    brightened. (b) The squashing factor $Q$ in logarithmic scale;
    samely contours are plotted at $\pm 1000$~G for photospheric
    $B_{z}$ and the dashed box outlines the flare location. (c) The
    $B_{z}$ map with the flare location outlined by a black box which
    is enlarged in panel (e). (d) Same as (b) but for the
    expansion-contraction degree $K$. (e) The vector magnetogram in
    the flare location and the thick white line denotes the bald patch
    (BP). Panel (f) gives some examples of the field lines (i.e., the
    BP separatrix field lines) that are tangent to the photosphere at
    the BP (the thick white line).}
  \label{fig:FlareBP}
\end{figure*}

\begin{figure*}[htbp]
  \centering
  \includegraphics[width=\textwidth]{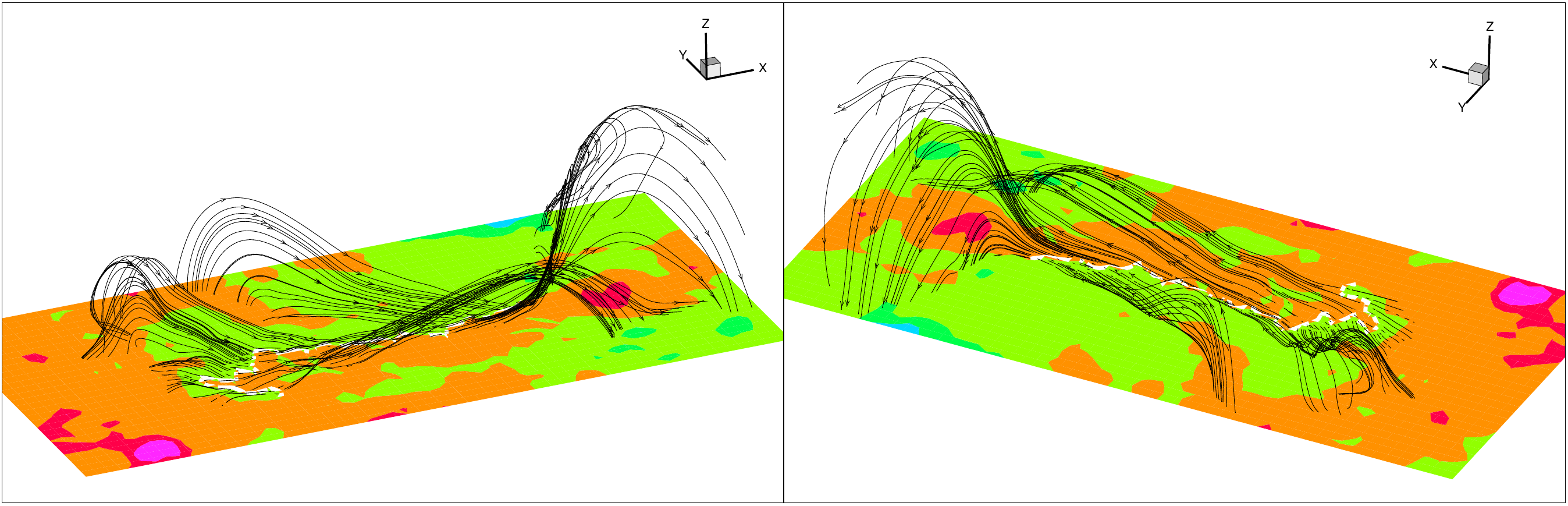}
  \caption{Different 3D views of the BP-separatrix field lines plotted
    in panel (f) of \Fig~\ref{fig:FlareBP}. The BP is denoted by the
    thick white lines. The $z$--axis scale is doubled for a better
    view of the field lines.}
  \label{fig:FlareBP3D}
\end{figure*}

Considering that the magnetic field is nearly steady with time, we
only compute the QSLs for a single frame. \Fig~\ref{fig:FlareBP}
depicts the $Q$ and $K$ maps (panel (b) and (d)) for the magnetic
field at 22:12 and compares with the AIA image at the same time. We
use a logarithmic scale since the squashing factor becomes abruptly
very large inside ths QSLs (e.g., \citet{Titov2002} defined the QSL as
a region with $Q \gg 2$). Note that there are data gaps in the maps
because the field lines there are opened, i.e., with ends of the lines
reaching the side or top boundaries of the computational volume. As
shown by the $Q$ and $K$ maps, the structures associated with data
abrupt change, i.e., the QSLs, are consistent between both maps. The
whole structure of the $Q$ map is rather complicated and may deserve a
comprehensive study, while here we put our focus on analyzing the
relation of the flare location (outlined by the dashed rectangle on
the AIA image) with the QSLs. Indeed, the flare location is clearly
co-spatial with the QSL of which the squashing factor reaches $\sim
10^{3}$ (see the region in the dashed rectangle in the $Q$ map). Then
why is this subregion has a strong change of magnetic connectivity? In
the same figure, we show the vector magnetogram (panel (e)) and the
field lines (panel (f)) in the same but a little larger subregion
outlined by the black rectangle in panel (c). The vector magnetogram
and the field lines reveal that underlying the flare region a bald
patch (BP) is located at the central portion of a long PIL (enhanced
by the thick white line in panels (e) and (f)). By its definition, BP
is a portion of PIL with $(\vec B\cdot \nabla)B_{z} > 0$, which means
the horizontal field at PIL crosses from the negative to positive
$B_{z}$ \citep{Titov1993,Bungey1996}, just oppossite to the normal
case. In the middle of the BP, the transverse field is nearly parallel
with the PIL, suggesting that it is not a single BP but fragments into
two parts. BP can also be defined as the locations where the magnetic
field is tangent to the photosphere, and the continuous set of field
lines that graze the surface at the BP form two separatrix surfaces
which separates three different topological regions. The separatrix
field lines are shown in \Fig~\ref{fig:FlareBP} (f) and with 3D views
in \Fig~\ref{fig:FlareBP3D}. Several studies have demonstrated that BP
can be correlated with flares and even CMEs due to the BP
separatrices, in which strong current sheets can be formed by
photospheric motions or flux emergence and trigger reconnection
\citep[e.g.,][]{Aulanier1998,Fletcher2001,Mandrini2002,WangTJ2002}. The
topological analysis of the flare site here, thus, suggests that the
AR 11117 flare can also be interpreted as a {\it bald-patch flare}
\citep{Aulanier1998,Delannee1999}, and may provide an evidence in
favor of reconnection along BP separatrices. The heights of the apexes
of the BP separatrices is about $2\sim 3$~Mm, meaning that the flare
happened rather low near the photosphere. However, in which way the
current sheet is formed and how the reconnection is triggered are not
clear and further study relying on higher-resolution and cadence data
could be needed. For another evidence of the presence of the BP
co-spatial with the flare, a curved dark feature near the flare
location, shown by the arrows on the AIA image (see panel (a) of
\Fig~\ref{fig:FlareBP}), has the same shape with the field lines near
the right end of the BP (indicated by the arrows in panel (f)). This
can be well explained by the dip of field lines just above the BP,
which can support dense cold plasma against gravity in the same way as
filaments.

\subsection{Energy and current}

In order to quantify the change of the field in the time series, we
computed a set of parameters and summarized them in
Table~\ref{tab:table1}. They include the total unsigned magnetic flux
$|\Phi|_{\rm tot} = \int_{S}|B_{z}|\rmd S$ (where $S$ represents the
photosphere), the total unsigned current $|I|_{\rm tot} =
\int_{S}|j_{z}|\rmd S$ (i.e., integral of the unsigned vertical
current on the photosphere), the total energy $E_{\rm tot} =
\frac{1}{8\pi}\int_{V}B^{2}dV$, the potential energy $E_{\rm pot}$,
the free energy $E_{\rm free} = E_{\rm tot}-E_{\rm pot}$, and the
ratio of free energy to potential energy. All these parameters are
important to characterize the evolution of the coronal magnetic
field. The first four parameters, i.e., the magnetic flux, the
current, the total and the potential energies, all keep increasing
with time in spite of the small flares. This is due to that the energy
injected by emerging magnetic flux is larger than the energy released
by the flares \citep[e.g.,][]{Regnier2005,He2011}. The total energy
and the potential energy are on the order of $10^{32}$~erg which is a
typical energy content of a medium-sized active region. Because of the
non-potentiality of the field, the total energy $E_{\rm tot}$ is
always higher than that of the potential field, which holds a energy
minimum state with a given magnetic flux on the photosphere. It is
commonly believed that the free magnetic energy plays a fundamental
role in flares, because the source of the energy for the flare must be
magnetic and only a fraction of the total magnetic energy, i.e., the
free energy can be converted to the kinetic energy and radiation of
flare \citep{Priest2002}. Our computation shows that the free energy
is on the order of $10^{31}$~erg (close to $10^{32}$~erg) which seems
to suffice to power a moderate flare, and this energy initially
increased like the total and potential energies before the C-class
flare started at 22:06 UT. One should bear in mind that even the free
energy is only partially involved with flares since the field after
flares is still non-potential and nonlinear \citep[e.g.,
see][]{Schrijver2009}. The energy released by the flare ought to be
quantified by the change in free energy from immediately before to
after the flare. Although the total energy increased even in the
interval of the flare, the free energy dropped as expected at 22:12,
i.e., the peak time of the flare, with a small amount of about
$1.7\times 10^{30}$~erg. It has been estimated that for the largest
flares up to X-class, the energy released are on the order of
$10^{32}$~erg \citep[e.g.,][]{Priest1981Book,Priest1987Book}. Thus by
a rough estimation, the decrease in the free energy of pre- and
post-flare is actually adequate to power this minor flare, of which
the energy needed is about several percents of the largest
class. Nevertheless, caution is needed when estimating the energy
budget of the flare by the drop of free energy in our modeling, since
many aspects of the model and the specific approach may influence the
results. We will discuss this in the conclusion section. In the last
column of Table~\ref{tab:table1} we calculated a mean vector deviation
between field $\vec b$ at each time with respect to the field $\vec B$
at time 21:00
\begin{equation}
  e_{\rm m} = \frac{1}{M}\sum_{i}\frac{|\vec b_{i}-\vec
    B_{i}|}{|\vec B_{i}|}
\end{equation}
where $i$ denotes the indices of all the pixels of the computational
volume and $M$ is the total number of the pixels. As a metric
monitoring the numerical variation of the field with the time, the
very low values of $e_{\rm m}$ again show that the change of the field
are rather small.

\begin{table*}[htbp]
  \centering
  \begin{tabular}{llllllll}
    \hline
    \hline
    Time & $|\Phi|_{\rm tot}$[$10^{22}$~Mx] & $|I|_{\rm tot}$[$10^{13}$~A] &
    $E_{\rm tot}$[$10^{32}$~erg] &  $E_{\rm pot}$[$10^{32}$~erg]
    & $E_{\rm free}$[$10^{31}$~erg] &  $E_{\rm free}/E_{\rm pot}$ &
    $e_{\rm m}$\\
    \hline
    21:00 &1.60 &  4.77 &  4.80 &  4.04 &  7.66 &  0.19 &0.00\\
    21:36 &1.63 &  4.80 &  4.95 &  4.18 &  7.69 &  0.18 &0.09\\
    22:00 &1.65 &  4.84 &  5.05 &  4.27 &  7.78 &  0.18 &0.11\\
    22:12 &1.66 &  4.92 &  5.09 &  4.33 &  7.61 &  0.18 &0.10\\
    22:36 &1.68 &  4.96 &  5.21 &  4.41 &  7.95 &  0.18 &0.11\\
    23:00 &1.70 &  5.05 &  5.28 &  4.50 &  7.79 &  0.17 &0.13\\
    \hline
  \end{tabular}
  \caption{Variation of parameters with evolution of the
    field, see text for details.}
  \label{tab:table1}
\end{table*}

\begin{figure*}[htbp]
  \centering
  \includegraphics[width=0.48\textwidth]{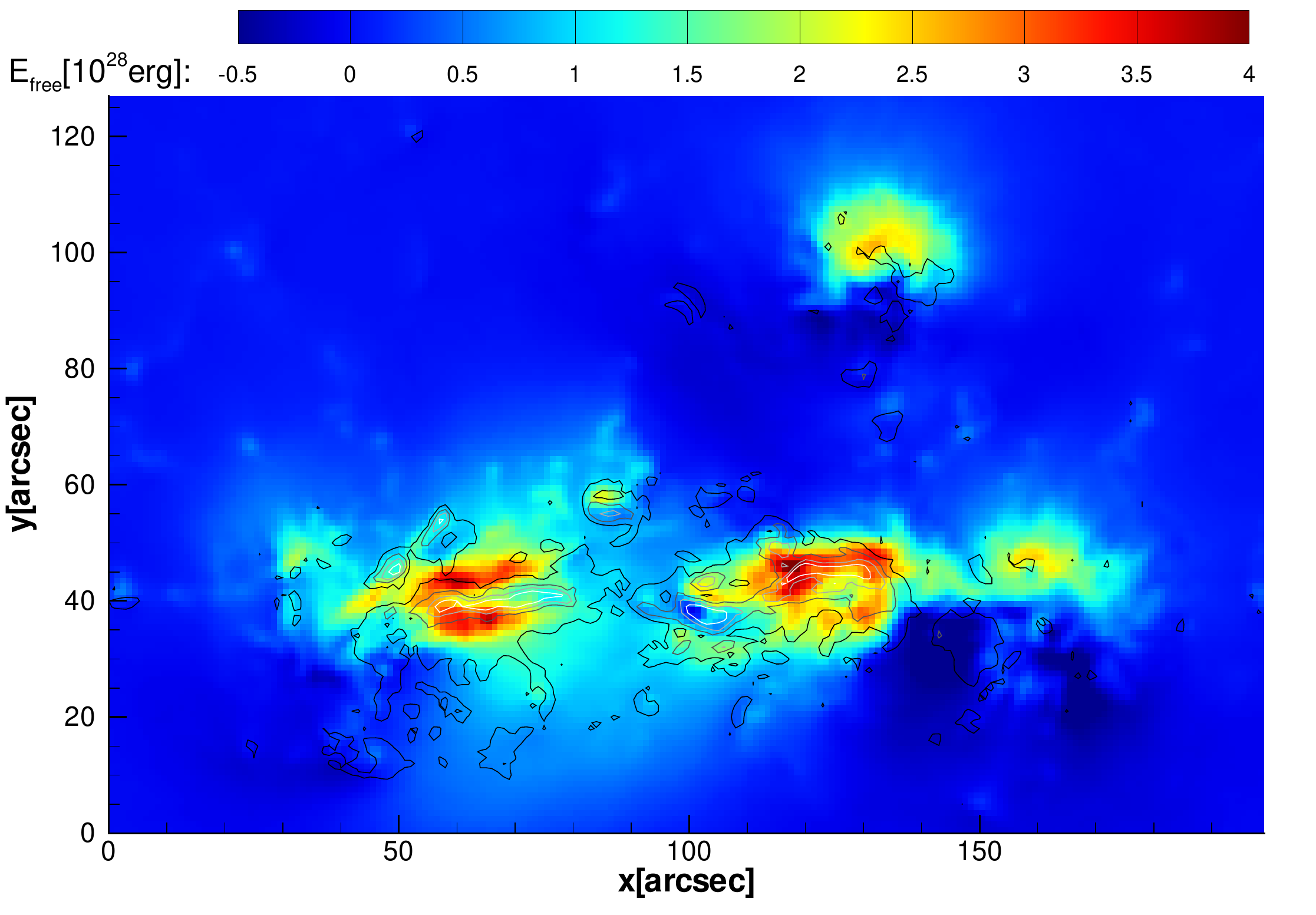}
  \includegraphics[width=0.48\textwidth]{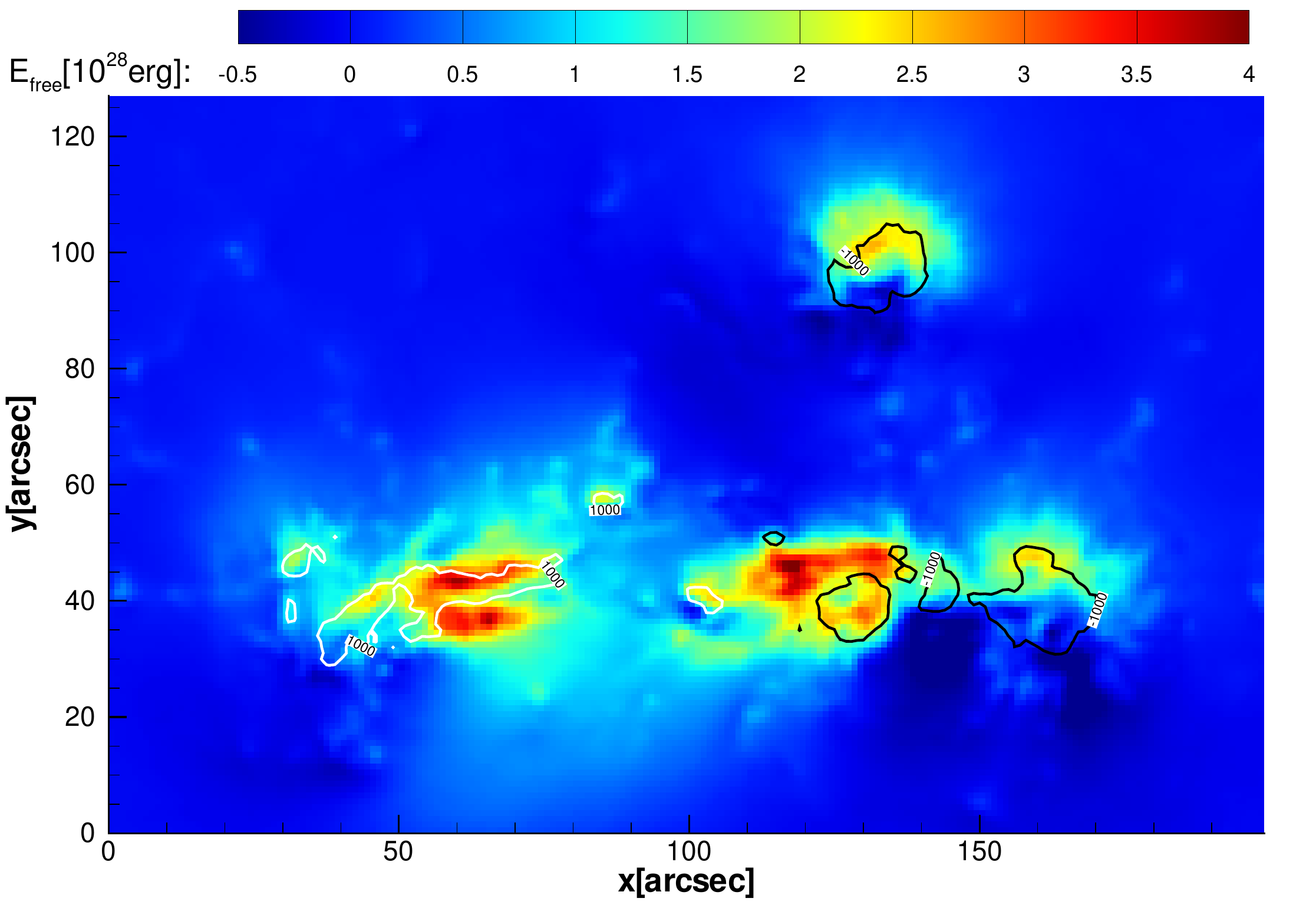}
  \caption{The images represent vertical integral of the free energy,
    showing the locations of free energy storage. The contour lines in
    the left panel represent the vertically integrated current density
    $\int |\vec J| \rmd z$ and the lines are color-coded with strength
    of the integrated current (increasing from black to white). The
    contour lines in the right panel represent $B_{z}$ on the
    photosphere with value of $\pm 1000$~G.}
  \label{fig:Efree_xy}
\end{figure*}

\begin{figure}[htbp]
  \centering
  \includegraphics[width=0.48\textwidth]{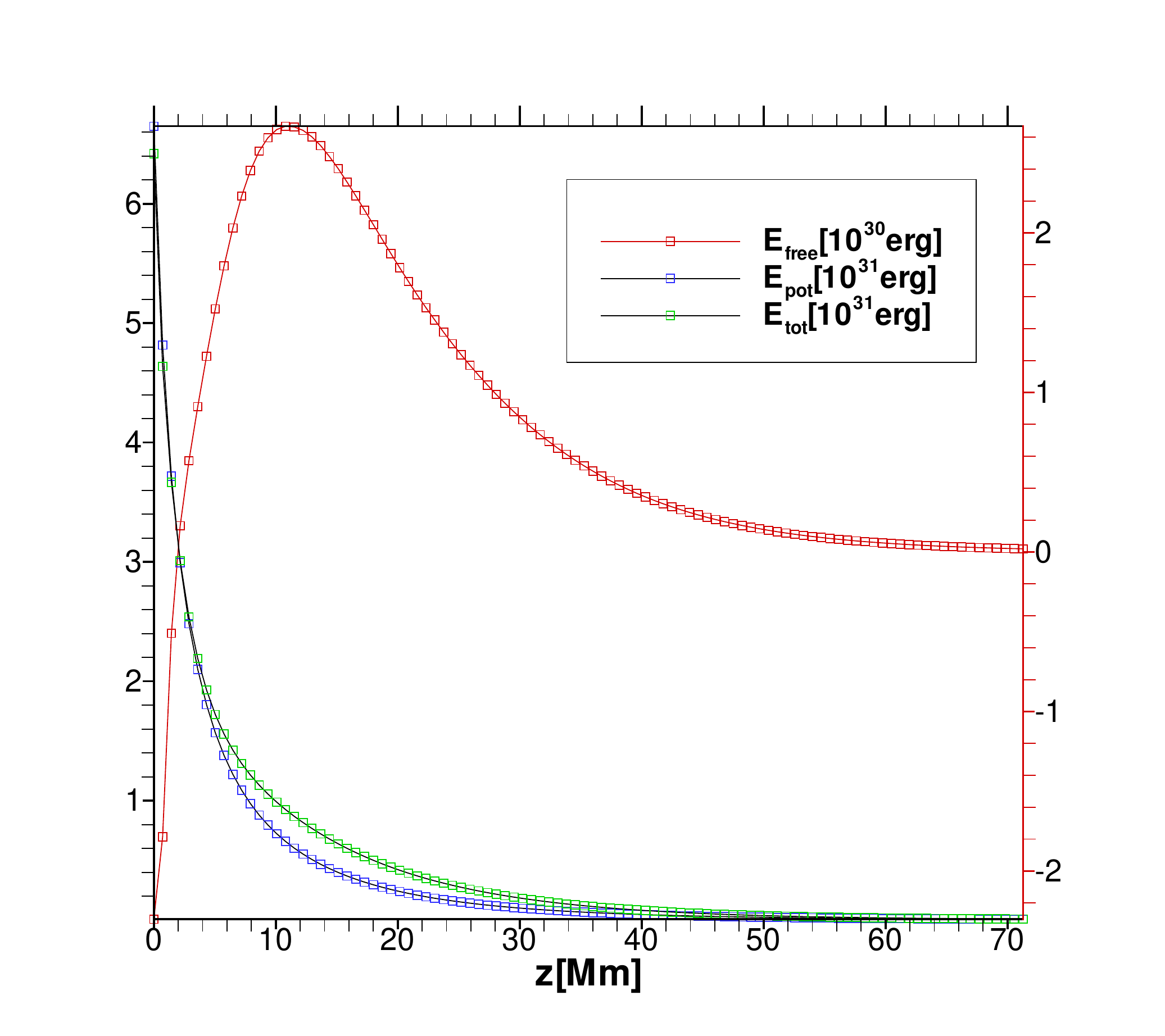}
  \caption{Variation of the horizontal surface integration of magnetic
    field energy along the $z$-axis. Note that the left vertical axis
    (black) indicates values for the $E_{\rm pot}$ and $E_{\rm tot}$
    and the right vertical axis (red) indicates values for $E_{\rm
      free}$.}
  \label{fig:Efree_zline}
\end{figure}

In addition to the global energy content, we can also study the
spatial distribution of the magnetic free energy, i.e., the locations
of the free energy storage. As an example, for the magnetic field at
time 22:00 we computed the vertical integration of the free energy
\begin{equation}
  E_{\rm free}(x,y) = A\int \frac{\vec B^{2}-\vec B_{\rm pot}^{2}}
  {8\pi}\rmd z
\end{equation}
where $ A = \rmd x\rmd y$, and plotted the distribution of $E_{\rm
  free}(x,y)$ on the horizontal plane (\Fig~\ref{fig:Efree_xy}). A sum
of the energy $E_{\rm free}(x,y)$ in the images gives the total free
energy listed in Table~\ref{tab:table1}. Over the left image of
\Fig~\ref{fig:Efree_xy}, the contour lines show the vertically
integrated current density $\int |\vec J| \rmd z$ and the lines are
color-coded with strength of the integrated current (increasing from
black to white); Over the right image the strongest regions of $B_{z}$
of photospheric field are outlined by the contour lines ($\pm
1000$~G). It can be clearly seen that the distribution of the free
energy is largely co-spatial with that of the current. This can be
easily understood because the coronal free energy (or the
non-potential energy) is actually stored in the current-carrying field
(where non-potentiality is strong). On the other hand, as shown by the
right image, the concentrations of free energy is not generally
spatially-correlated with those of strongest magnetic flux. It should
be noted that in the image there are some places with negative values
of the vertically integrated free energy. This is physically valid
since the there is no restriction that the energy density (and thus
any sub-volume energy) must always be greater than that of the
potential field, although a non-potential field must have a global
energy content greater than the potential field
\citep[e.g.,][]{Mackay2011}. In \Fig~\ref{fig:Efree_zline} we plot the
horizontal surface-integral of the total energy, the potential energy,
and the free energy, e.g.,
\begin{equation}
  E_{\rm free}(z) = \rmd z\int\frac{\vec B^{2}-\vec B_{\rm pot}^{2}}
  {8\pi}\rmd x\rmd y
\end{equation}
as functions of the height $z$. The total and the potential energies
are predominantly located near the photosphere where the magnetic
field strength is high, whereas the free energy (the red curve) is
situated mainly above the photosphere in a range of 5~Mm to 30~Mm with
its maximum at about 10~Mm. It is interesting to find that near the
photosphere, the free energy is negative with a minimum value at the
photosphere, which says that the observed vector field has a lower
surface energy content than the potential field. This is, however, not
surprising as we have noted that any sub-volume energy content of the
non-potential field may be lower than potential energy.

\begin{figure*}[htbp]
  \centering
  \includegraphics[width=\textwidth]{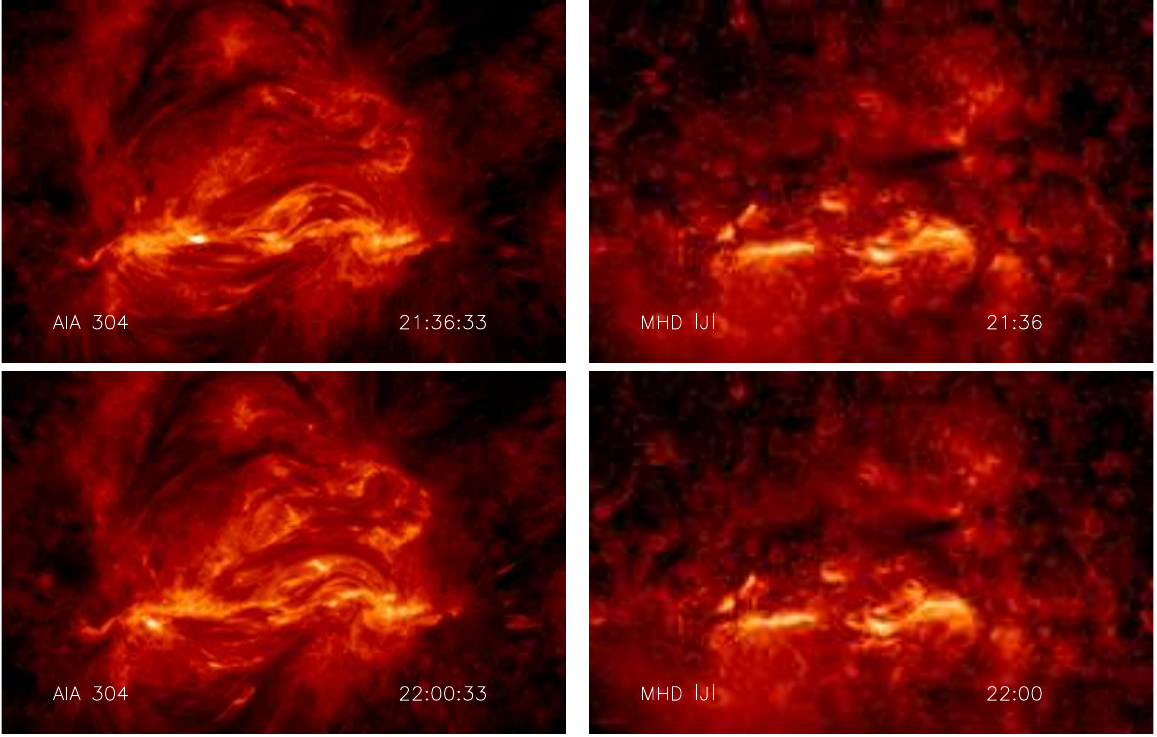}
  \caption{Left column is the AIA 304 {\AA} image and right column is
    the synthetic images of current using vertical integral of $J^{2}$
    computed by the MHD model.}
  \label{fig:AIA_304}
\end{figure*}

Electric current can characterize the non-potentiality of the field,
e.g., the patterns of strong current concentration may serve as a
proxy for the non-potential structures (e.g., the sigmoids) in the
corona \citep{Schrijver2008,Archontis2009,Sun2012}. In particularly,
the current structures are regions where reconnection may happen and
magnetic energy is converted to thermal energy and heating, and thus
creating hot emission. In \Fig~\ref{fig:AIA_304} we give examples of
the synthetic images of the current which is computed by vertical
integration of $J^{2}$ (i.e., $\int_{z}J^{2} \rmd z$, see
\citet{Archontis2009}) and compared with the AIA 304 {\AA}
images. Since the term $J^{2}$ is proportional to the Joule heating
term, thus it simulates but very roughly the hot emission. As can be
seen in the figure, the strong current regions are indeed coincident
with the regions with high intensity of emission. However, the result
does not show any intensive current associated with the flare
site. This may be because the current sheet in the BP separatrices is
very thin and is failed to be resolved by the present grid resolution.

\section{Conclusions}
\label{sec:conclude}

In this work, we have applied the data-driven CESE--MHD model to
investigate the 3D magnetic field of AR 11117 around the time of a
C-class confined flare occurred on 2010 October 25. Similar to the
field extrapolation method our model is designed to focus on the
magnetic field, but its nonlinear dynamic interactions with plasma and
finite gas pressure (denoted by plasma $\beta$) are also embedded,
although simplified. Assuming that the dynamic evolution of the
coronal magnetic field can be approximated by successive equilibria,
we have solved a time sequence of MHD equilibria basing on a set of
vector magnetograms for AR 11117 taken by {\it SDO}/HMI around the
time of flare. By analyzing the computed 3D magnetic field along with
the observation, we have the following results:

\begin{enumerate}
\item The model has qualitatively reproduced the basic structures of
  the magnetic field, as supported by the visual similarity between
  the field lines and the {\it SDO}/AIA loops, which shows that the
  coronal field can indeed be well characterized by the MHD
  equilibrium in most time. The magnetic field is very non-pontential
  with strong shear locally and some twists compared with the
  potential model. There are also some loops failed to be recovered by
  the MHD model, but only at the time set very near the happening of
  the flare. This means that the magnetic field is rather dynamic when
  the energy is suddenly released in a time-scale far shorter than
  that of relaxation by Alfv\'en speed.
\item The magnetic configuration changes very limited during the
  studied time interval of two hours, and the flare-related
  reconnection takes place at rather small scale and low height near
  the photosphere. The topological analysis reveals that the small
  flare is correlated with a BP and the energy dissipation can be
  understood by the reconnection associated with the BP
  separatrices. However, no intensive current is found in the flare
  site related with the BP separatrices. This may be because the
  current sheet associated with the separatrices is very thin and
  cannot be resolved by the present grid resolution. Further study
  exploiting the full resolution and high-cadense observations is
  required for explaining how the BP-flare is actived, e.g., where the
  current sheet is formed and how the reconnection is triggered.
\item Because of the continuous flux emergence, the total unsigned
  magnetic flux and current through the photosphere keep increasing
  (but very slightly) in spite of the flare. Although evolution of the
  total magnetic energy also exhibits the same tendency as that of the
  total magnetic flux, the sum of free energy for the computational
  volume drops when the flare happened, indicating that some of the
  non-potential energy is released by the flare. Our computation shows
  that the amount of the free energy loss is on the order of
  $10^{30}$~erg, which is adequate to power a minor C-class flare.
\end{enumerate}

In summary, our model capture the basic features of the 3D magnetic
field of the target active region both qualitatively and
quantitatively, also the results give some hints on the trigger
mechanism of the flare. Nevertheless, we remind the readers that the
results, especially in the quantitative aspect, should be interpreted
with caution because they can be influenced by many uncertainties in
the modeling. The uncertainties also exist in other models of the
similar kind, for example, the NLFF modeling, and even for the same
model, different codes may produce very inconsistent results
\citep{Schrijver2008,Derosa2009}. The uncertainties may first come
from measurement error of the HMI magnetogram data. For example,
\citet{Sun2012} have estimated that the free energy content could be
affected with several percents by the spectropolarimetric noise in the
magnetogram; even such small error is large for the present case in
which the flare may only release a very small fraction of the free
energy. It should be noted that in the NLFF model the systematic error
can be greater because of the force-free assumption and the
preprocessing and smoothing of the original data. Although our model
does not suffer from such preprocessed-related problem, the systematic
uncertainties can still come from the simplified configuration of the
solar atmosphere, the boundary conditions and the data interpolation
from the original non-uniform grid to a uniform grid when computing
the parameters. Especially the using of a low-$\beta$ plasma globally
is far from the realistic case in which the solar atmosphere is highly
stratified with much larger gas pressure near the
photosphere. Furthermore the assumption of static state of the
magnetic field is unjustified by the onset of the flare, which can
drive the field lines very dynamic and make our computation
unreliable, as discussed in the comparison of the MHD results with the
AIA images. This is a much more basic problem (than the others
aforementioned) encountered by any extrapolation of magnetic field
with static or quasi-static models.

Future improvements merit to be made in several aspects. To increase
the capability of adaptive resolving the small-scale structures can be
realized hopefully by the aid of the adaptive-mesh-refinement
technique. Exploiting more observations like the surface flows
computed by the LCT-type methods can further constraint the model and
provide important information for the realistic dynamic evolution of
the magnetic field. More physics-based thermodynamic model for the
solar atmosphere with stratified temperature will also be considered
to couple the photosphere and corona, in order to model the behavior
of the magnetic field in a highly stratified and inhomogeneous plasma
with $\beta$ from $>1$ to $\ll 1$.

\acknowledgments

The work is jointly supported by the 973 program under grant
2012CB825601, the Chinese Academy of Sciences (KZZD-EW-01-4), the
National Natural Science Foundation of China (41031066, 40921063,
40890162, and 41074122), and the Specialized Research Fund for State
Key Laboratories. Data are courtesy of NASA/SDO and the AIA and HMI
science teams. Special thanks go to our anonymous reviewer for
valuable suggestions for the improvement of the paper.



\end{CJK*}
\end{document}